# The FAIR Funder:
# A pilot programme to make it easy for funders to require and for grantees to produce FAIR Data

FAIR Funders Collaboration[1]

February 15, 2019

There is a growing acknowledgement in the scientific community of the importance of making experimental data machine findable, accessible, interoperable, and reusable (FAIR). Recognizing that high quality metadata are essential to make datasets FAIR, members of the GO FAIR Initiative and the Research Data Alliance (RDA) have initiated a series of workshops to encourage the creation of Metadata for Machines (M4M), enabling any self-identified stakeholder to define and promote the reuse of standardized, comprehensive machine-actionable metadata. The funders of scientific research recognize that they have an important role to play in ensuring that experimental results are FAIR, and that high quality metadata and careful planning for FAIR data stewardship are central to these goals.  We describe the outcome of a recent M4M workshop that has led to a pilot programme involving two national science funders, the Health Research Board of Ireland (HRB) and the Netherlands Organisation for Health Research and Development (ZonMW). These funding organizations will explore new technologies to define at the time that a request for proposals is issued the minimal set of machine-actionable metadata that they would like investigators to use to annotate their datasets, to enable investigators to create such metadata to help make their data FAIR, and to develop data-stewardship plans that ensure that experimental data will be managed appropriately abiding by the FAIR principles. The FAIR Funders design envisions a data-management workflow having seven essential stages, where solution providers are openly invited to participate. The initial pilot programme will launch using existing computer-based tools of those who attended the M4M Workshop.

## Introduction

Following a recent "Metadata for Machines" Workshop at the GO FAIR Offices in Leiden[2], a number of international stakeholders (including two national funding agencies) conceived a pilot programme to demonstrate how increasingly FAIR research outputs can be realistically mandated by funding agencies. The Pilot attempts to address two fundamental difficulties faced by funders that wish to require FAIR Data Stewardship as part of supported research projects: (1) assessing the quality of the overall data stewardship plan for the project and (2) assessing the level of FAIRness of research outputs. Both of these assessment activities require deep technical know-how and professional experience that is currently in short supply and is not typically found within public or private funding organizations. The Pilot will allow us to study the mechanisms whereby funding agencies can explicitly define expectations with respect to the content of the metadata that investigators, when responding to the funding calls, will use to make their data FAIR. Together, the participants of this Pilot aim to demonstrate that funding agencies can realistically and routinely **require** machine-actionable metadata as part of funded projects, while grantees can realistically and routinely **comply** without any additional training or specialised knowledge about metadata. This is fundamental to the overall vision of an Internet of FAIR Data and Services.

---

[1] Appendix A
[2] M4M workshop series: https://www.go-fair.org/resources/go-fair-workshop-series/ ; Workshop documents;
The first M4M Workshop was held in Leiden, October 15-16 2018, *Metadata 4 machines help you find and (re)use relevant research data*, Kristina Hettne (November 2, 2018)
https://digitalscholarshipleiden.nl/articles/metadata-4-machines-help-you-find-and-reuse-relevant-research-data

The Pilot describes seven basic functional stages and features possible solution providers (designated by their logos in the diagram) servicing each stage. The list of current participants in this Pilot reflects only the enthusiasm of early movers who were in attendance at the Metadata for Machines Workshop[1], and should not be interpreted as being the only or best configuration of tools and organizations to realize a FAIR Funders ecosystem. To the contrary, the Pilot is conceived to be the first in an ongoing programme, permitting what is anticipated to be a large number of potential stakeholders to voluntarily join when they are able. This open and extensible mechanism by which others may join, without disturbing the current work, will be developed in subsequent publications.

By presenting the FAIR Funders Pilot in this way, we intend to showcase its feasibility. The primary focus on machine-actionable metadata is embedded in (and powered by) the context of FAIR data stewardship in general. As such, the Pilot utilizes a highly customisable data stewardship planning tool (the Data Stewardship Wizard, or DS Wizard[3]) as a platform to guide researchers in the use and reuse of machine-actionable metadata standards. The process by which machine-actionable metadata templates are created and deployed is discipline agnostic, and entirely generic, and can be easily replicated in future funding calls for other topics in any research domain.

---

[3] https://ds-wizard.org

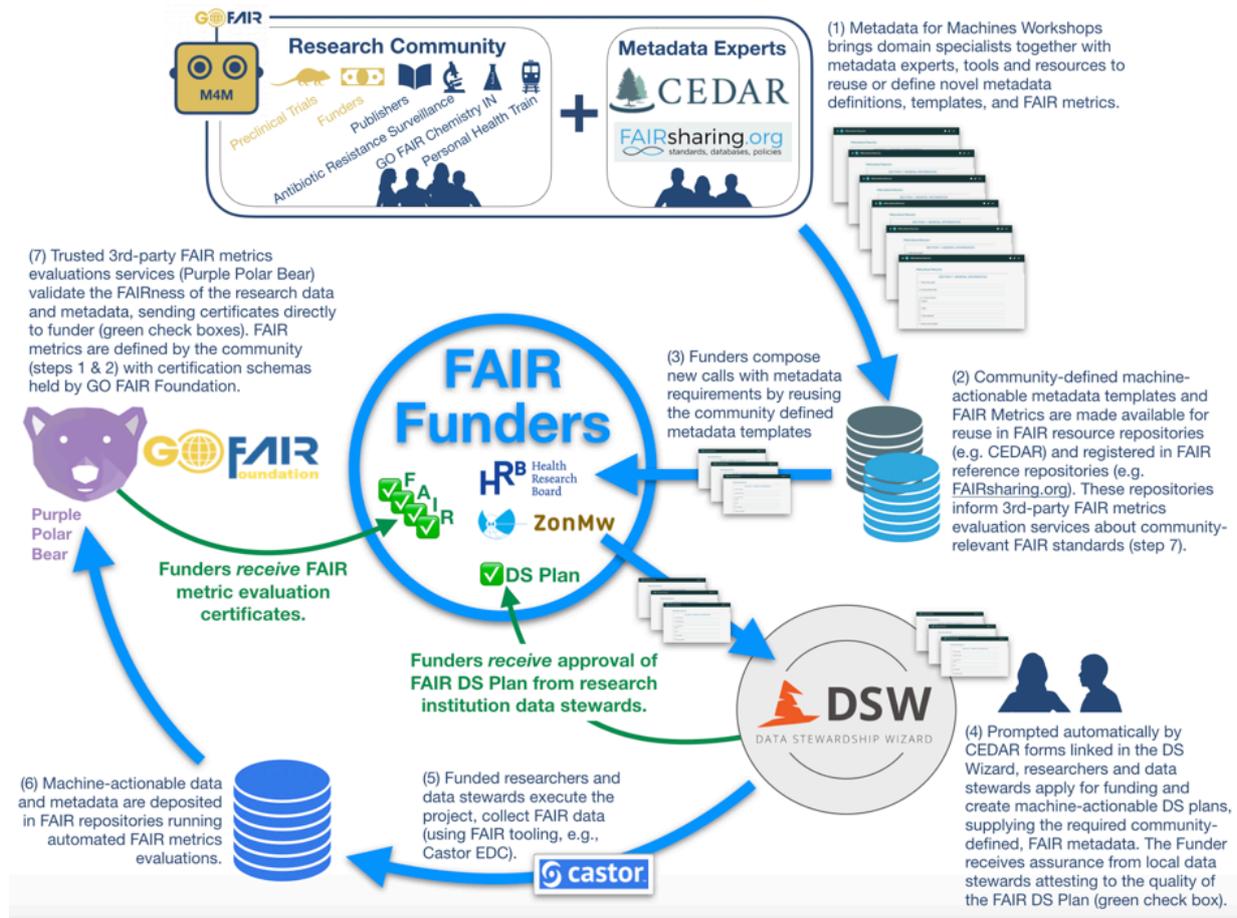

Figure 1: FAIR Funders workflow incorporating seven essential stages necessary to complete a funding cycle that supports the creation and sharing of FAIR research outputs. Research funders are located in the middle of the diagram (blue circle) while other key stakeholders in the process produce FAIR data or metadata, or provide supporting services. The green arrows indicate emerging solutions to two fundamental problems faced by funders today, that want FAIR Data Stewardship in their funded research projects: (1) Assessing the quality of the overall data stewardship plan for the project, and (2) assessing the level of FAIRness of research outputs. In each case, the process proposed here removes the responsibility of the technical assessments from the funder, and places it with the appropriate stakeholders of the research community that have the expertise to make meaningful assessments. These external assessment services require a combination of expert training (data stewards), content (FAIRsharing) technology (CEDAR; DS Wizard; Castor EDC; FAIR Data Points; FAIR Metrics Evaluators), and trusted third-party certification organisations (GO FAIR Foundation; Purple Polar Bear software development). Although there is an agreement in the broad outline of this workflow, the organisation of stakeholders and the technical specifications linking the services will be established in subsequent planning. Not pictured in this high-level view of the workflow is information flow between the components that ensure standards are always updated and shared in common.

# The FAIR Funders Pilot in more detail

### *Stage (1) Research Communities run Metadata for Machine Workshops.*

Metadata for Machine (M4M) workshops[4] are extremely popular, ongoing events that bring metadata experts together with domain specialists (1) to create community-defined, FAIR-compliant[5] metadata schemas and FAIR metrics[6], (2) to encode these schemas as machine-actionable templates and (3) to make these templates available for reuse to the larger community via open registration services and repositories.

M4M Workshops were developed realizing that specialized metadata for the F and R Principles can in many cases, be created only by the practicing experts in those domains, and yet these experts typically have limited (if any) working knowledge of metadata. Although training is part of the goals of the M4M workshop, these events are intended to be an efficient, fast-track mechanism to the creation of practical (though not necessarily pristine) FAIR metadata content[7]. Although a number of applications exists for metadata creation and sharing, this Pilot will utilize the semantically enabled metadata capture tooling from the Center for Expanded Data Annotation and Retrieval (CEDAR)[8] and the community-supported[9] FAIRsharing[10] resource for the registration of standards (metadata, identifiers schemas and FAIR metrics), repositories and data policies and will also connect to the FAIR evaluator and the DS Wizard.

As coordinators who have overarching perspective regarding related metadata efforts in a variety of scientific domains, GO FAIR and RDA play a special role as metadata experts in M4M Workshops, helping to maximise the efficient reuse of metadata templates and associated ontologies among research communities that are often siloed, and do not exchange this information[11]. GO FAIR and RDA can thus accelerate the reuse of mature curated templates while preventing wasteful re-invention of previous work that often frustrates interoperation.

Pictured in Figure 1 are five examples of self-identified research communities that have either completed, or are planning M4M workshops. The FAIR Funders Pilot programme was launched (January 14, 2019) with a use case from the Preclinical Trials community[12]. The outputs of the Preclinical Trials M4M workshop were agreed-upon metadata elements (such as research project registration forms that are to be completed before data collection begins) and their corresponding machine-actionable templates. The metadata templates can be exposed as automatically generated webforms using CEDAR (webforms are represented in Figure 1 as white Webpage icons with blue headers). CEDAR allows practicing researchers (or data stewards) to routinely "build" FAIR machine-actionable metadata with minimal training.

---

[4] M4M #1 https://www.go-fair.org/resources/go-fair-workshop-series/metadata-for-machines-workshops/ ; https://osf.io/qe9fa/

[5] Wilkinson, M. D. *et al.* The FAIR Guiding Principles for scientific data management and stewardship. *Sci. Data* 3:160018 doi: 10.1038/sdata.2016.18 (2016) and https://www.go-fair.org/fair-principles/

[6] Wilkinson, M. D. *et al*. A design framework and exemplar metrics for FAIRness. *Sci. Data* 5:180118 doi: 10.1038/sdata.2018.118 (2018) and http://fairmetrics.org

[7] The M4M approach is inspired by the mantra used by developers of early computer networking systems of "rough consensus and working code", see Strawn G.O., Chapter 83: From ARPAnet, through NSFnet, to Internet, In: Leadership in Science and Technology: A Reference Handbook, Edited by: William Sims Bainbridge. DOI: http://dx.doi.org/10.4135/9781412994231.n83

[8] https://metadatacenter.org

[9] Sansone S.A. et al. FAIRsharing, a cohesive community approach to the growth in standards, repositories and policies. *Nat Biotech* (*accepted*), pre-print: https://doi.org/10.1101/245183

[10] https://fairsharing.org

[11] The metadata groups of RDA have been trying to understand different domain approaches. There is a catalog of metadata schemas organised by MSCWG https://www.rd-alliance.org/groups/metadata-standards-catalog-working-group.html (original directory is at http://rd-alliance.github.io/metadata-directory/ and the more recent catalog at https://rdamsc.dcc.ac.uk/). The 'umbrella' MIG https://rd-alliance.org/groups/metadata-ig.html has a set of metadata principles and is working on a canonical set of metadata elements (with the idea that common metadata formats can use it as the interconversion medium).

As key stakeholders, the Health Research Board of Ireland (HRB) and the Netherlands Organisation for Health Research and Development (ZonMW) have also run a joint "Science Funders" M4M workshop[13]. In this workshop, science funders, researchers, and data stewards worked together to create the agreed-upon FAIR metadata descriptions that are critical to sponsored research. These elements include unique and persistent identifiers for referencing funding announcements and research calls for proposals, and for long-term tracking of research outputs. CEDAR forms were built to capture all essential grant information, including documentation of resulting publications and data assets. As part of this Pilot, all subsequent publications and data assets produced in the course of funded research project can be linked to the completed forms by filling in the uniquely identified references (i.e., machine-actionable metadata).

Future M4M Workshop topics have been proposed for academic publishers, antimicrobial resistance surveillance networks, Integrating the Healthcare Enterprise, the GO FAIR Chemistry Implementation Network and the Personal Health Train initiative.

### *Stage (2) Community-defined metadata templates & FAIR metrics are stored in Open repositories.*

The technical outputs of M4M workshops are cumulative, searchable and shared, leading to rapid and widespread reuse of FAIR metadata and FAIR metrics within the many different research domains. Information about Preclinical Trials metadata standards will be listed in FAIRsharing.org, a curated resource on data and metadata standards, databases, and data policies. At FAIRsharing.org, stakeholders are guided in the discovery, selection and re-use of metadata standards and FAIR metrics. In the case of community-defined FAIR metrics, web services, such as those that provide trusted third-party FAIR metrics evaluation (Stage 7), can access the community standards that are necessary for those metric evaluations. FAIRsharing is also working to create machine-readable metadata standards that can be combined to: (i) create metadata templates to ensure that a dataset is described according to one or more relevant community-defined metadata requirements (e.g. a biodiversity experiment is described according to Darwin Core standards[14]) and (ii) be used by the FAIR evaluator to evaluate the compliance of a dataset against one or more community-defined metadata requirements.

### *Stage (3) Funders use the CEDAR repository of community-defined metadata templates and FAIR metrics to compose new call requirements, and to embed them in the DS Wizard Knowledge Model. FAIRsharing provides the guidance to the community-defined metadata standards. Compliance with FAIR Principles is reported at the end of the project by automated services (Stages 6 & 7).*

---

[12] Preclinical Trials is a term of art describing any research involving animal (but non-human) subjects. See https://preclinicaltrials.eu ; M4M #2 https://www.go-fair.org/events/m4m-2-preclinical-trials-m4m-3-funders/ ; https://osf.io/924md/
[13] M4M #3  https://www.go-fair.org/events/m4m-2-preclinical-trials-m4m-3-funders/ ; https://osf.io/h2mpr/
[14] https://doi.org/10.25504/FAIRsharing.xvf5y3

Community-driven metadata standards are essential to drive harmonization, achieve interoperation and maximize reuse of data.  It is pivotal to guide the data producers to capture essential metadata elements from the onset, rather than upon submission to a repository, find out that some metadata field is missing, or that a specific controlled vocabulary was required. By tracking which metadata standards have been defined by a given community or disciplines (e.g. in the agro community https://fairsharing.org/collection/AgBioData) and which repositories require which metadata upon data deposition, FAIRsharing provides guidance on which community-defined metadata elements need to be used by the researchers when describing their datasets. At the same time, the accumulated machine-actionable templates stored in CEDAR constitute a growing pool of community-defined FAIR metadata elements and FAIR metrics that can be consulted when anyone wishes to compose or capture machine-actionable metadata profiles.

In the case of science funders, program managers can easily discover, and then compose FAIR standards-compliant metadata requirements and metrics as part of the requirements for submission of research proposals and/or as part of FAIR data stewardship plans without any specialised skills (for example, via 'drag and drop' interfaces allowing program managers to compose, test and refine metadata profiles).

In the case of applicants and grantees, the selected FAIR standards-compliant metadata templates will be exposed as corresponding webforms to be completed by the investigators during the grant application process and in FAIR data management planning. To the investigator and institute managers, the CEDAR forms look similar to the familiar webforms that are completed as part of ordinary application processes. However, CEDAR forms are able to prompt and capture the desired metadata in machine-actionable format (for example, using autocomplete functions and controlled vocabularies to help the researcher speed entry and provide unambiguous controlled terms for concepts like people, affiliations, organism names, disease phenotypes, chemical names, geolocation, soil types, and others).

In this way, funders reach into a growing pool of community-defined metadata templates and FAIR metrics, and compose domain-appropriate, machine-actionable metadata profiles that also match the requirement of the target repositories. Then, investigators complete the corresponding webforms, creating FAIR, machine-actionable metadata descriptions (themselves with unique and persistent identifiers) that will be permanently linked to each research project.

ZonMw and HRB have been early movers in FAIR data management and data stewardship planning. This work has included earlier pilots that have evaluated FAIR data stewardship plans and mechanisms for training data stewards with FAIR skill sets. Based on this rich experience, both organisations have recently decided in this FAIR Funders Pilot to delegate the planning and evaluation of FAIR data stewardship to research organizations, as is usually done for compliance with ethics policies in human and animal research[15]. In these instances, the funder requires only that local

---

[15] In 2019 ZonMw is gradually introducing a new procedure for data management in the projects it funds by: (1) requiring a minimal set of key elements for making data reusable; (2) promoting the use of standards within research communities for interoperability; (3) putting research institutes and their data stewards in the lead to work with grantees to deliver a good quality DMP and to meet the requirements; (4) evaluating the outcome at the end of a project according to the required items. The development and introduction is done in collaboration with data professionals from research institutes and data services.  ZonMw is starting to involve other health funding agencies in the Netherlands as well. At present, ZonMw is implementing the procedure in existing systems. In 2019, ZonMw will start to organise small scale pilots in funding calls to test elements of the funder pilot. These are aimed at making the workflow eventually machine actionable.  In June 2018, HRB co-funded FAIR training for 13 data stewards from 7 Irish research organisations. In October, the HRB launched two funding schemes aiming to

ethical committees have reviewed and attest to satisfactory treatment of human and animal research subjects. In a similar way, in this Pilot ZonMw and HRB will require FAIR data stewardship, but they will leave the creation and the evaluation of the FAIR data stewardship plan to trained (and perhaps one day certified) data stewards working at the research organisations. Funding of research projects will require the data stewards' assurance that compliance with best practices and with community standards in FAIRsharing and FAIR data stewardship is in place. Furthermore, ZonMw and HRB have also decided in this Pilot to monitor the level of FAIR compliance achieved in the course of the research project using trusted, third-party, automated FAIR evaluation services as professionalized extensions of existing prototypes[16].

### *Stage (4) Researchers & Data Stewards apply for funding and write data-stewardship plans, supplying the required community-defined metadata (prompted automatically by CEDAR forms linked in the DS Wizard and FAIRsharing).*

The previous stages were intended to 'create' a pool of machine-actionable metadata necessary to power FAIR applications. Once created, these templates can be used and reused by the community to make metadata capture as easy and efficient as possible. Although many data management and data stewardship planning tools currently exist, this Pilot will deploy the Data Stewardship Wizard[17]. The DS Wizard is a joint project initially launched by ELIXIR NL (Dutch Techcentre for Life Sciences) and ELIXIR CZ (Czech Technical University and Institute of Organic Chemistry and Biochemistry AS CR) for the life sciences domains. It is a smart data management planning tool for FAIR Open Science based on a hierarchical machine-actionable knowledge model (KM) regarding good practices in data stewardship[18], mainly in the form of questions and answers (grey circle in Figure 1).

The DS Wizard KM is openly available[19] and is highly customizable. For example, a GO FAIR-sponsored hackathon in July 2018 brought developers together in Leiden for 3 days, to extend the KM with quantitative metrics for FAIRness, Openness and Good Practice in data management[20]. The KM can also be easily extended with relevant community-specific information from FAIRsharing, FAIR metrics and appropriate CEDAR metadata webforms. This ensures that the users are informed and guided to compliance with the relevant generic, as well as domain-specific metadata standards.

In a similar way, in our Pilot programme, ZonMw and HRB will use the results of the M4M workshops to have the generic DS Wizard KM extended with the appropriate CEDAR webforms, FAIR metrics, and links to any appropriate set of FAIRsharing records. ZonMw and HRB will then encourage grant

---

fund up to 30-35 projects 30 research calls, where FAIR data stewardship plans will be explicitly eligible for funding supporting (on average) 5% of project funding. These funded research projects will complete an entire funding cycle by 2021 and will be the first test of funded, FAIR data stewardship plans in action.

[16] Academic prototype FAIR Metric Evaluator http://linkeddata.systems:3000/FAIR_Evaluator/ ; Professional prototype FAIR Metric Evaluator http://www.bio-itworld.com/2018/06/06/finding-the-usable-in-fair-data-at-bioit18.aspx

[17] "Data Stewardship Wizard": A Tool Bringing Together Researchers, Data Stewards, and Data Experts around Data Management Planning. Robert Pergl, Rob Hooft, Marek Suchánek, Vojtěch Knaisl, and Jan Slifka. Submitted for publication in Data Science Journal.

[18] More on machine-actionable DS Plans: Machine actionable DMP's at IDCC in 2017 https://doi.org/10.3897/rio.3.e13086 ; Ten principles for machine-actionable data management plans https://zenodo.org/record/1461713#.XGZxXS2ZNKQ ; Donnelly M., Jones S., Pattenden-Fail J.W. (2010) DMP Online: A Demonstration of the Digital Curation Centre's Web-Based Tool for Creating, Maintaining and Exporting Data Management Plans. In: Lalmas M., Jose J., Rauber A., Sebastiani F., Frommholz I. (eds) Research and Advanced Technology for Digital Libraries. ECDL 2010. Lecture Notes in Computer Science, vol 6273. Springer, Berlin, Heidelberg.

[19] Data Stewardship Wizard Knowledge Model: https://github.com/ds-wizard/ds-km

[20] The evaluation of the metrics is based on G,O,F,A,I,R measures assigned to each of the 600+ questions composing the current Wizard KM. The metrics are then computed as a weighted sum.

applicants to use the augmented DS Wizard to create FAIR machine-actionable community-defined metadata as part of their FAIR data stewardship plans. The funders will not actively review the resulting plans, but instead will task (certified) data stewards at local research organisations to assert that a satisfactory plan is in place. This comes in the form of "report assemblers" -- a functionality to put together a condensed report of most important information required by funders. This is represented by various templates, such as the one based on Science Europe recommendations[21]. The output of the DS Wizard is both a human-readable but also a machine-actionable, FAIR, Science Europe complaint, Data Stewardship Plan.

### *Stage (5) Funded researchers execute their projects, collecting experimental data.*

Researchers committed to FAIR data have a growing tool kit in which to draw from, to facilitate increasingly FAIR data capture. Both CEDAR and the Castor EDC electronic case-report form system[22] will be used to capture FAIR machine-actionable metadata and data throughout the project. Because CEDAR templates can be created to satisfy or validate many of the FAIR principles as users enter their metadata, the metadata-entry process will provide immediate feedback that will help make the metadata more FAIR, including indications of what steps must be be taken to improve the metadata. Castor EDC Forms and Form Templates can be annotated with metadata, including automatically consuming metadata produced within CEDAR and the DS Wizard. The data and metadata collected within the platform can be made available in the Castor FAIR Data Point.

CEDAR, the DS Wizard, and Castor EDC will pursue interoperability across their platforms. In some cases, these systems may also provide access to external FAIR metadata evaluators, so that external FAIR evaluation services can provide real-time assessments of ongoing metadata updates (Stage 7).

### *Stage (6) Machine-actionable data and metadata are deposited in repositories running automated FAIR metrics evaluations.*

At the conclusion of the research project, data and metadata will be deposited in one or more repositories supporting FAIR data (such as FAIR Data Points[23]). Data collected and annotated in Castor EDC can be made accessible via the Castor FAIR Data Point. In this Pilot, the FAIR repository will be augmented with functions that automatically trigger third-party FAIR metric evaluation services.

### *Stage (7) Trusted third-party FAIR metrics evaluation service.*

Supporting repositories and FAIR Data Points will expose metadata necessary for FAIR Metrics evaluation[24]. By request of and support by the Dutch Ministry of Economic Affairs, the GO FAIR Foundation[25] (GFF) will concentrate in 2019/2020 on the definition, design and implementation plan for a coherent certification program for a number of GO FAIR related services and products. Certifications are foreseen for datasets, meta-datasets, FAIR Data Points, individual FAIR practitioners, service suppliers, organisations / institutions and ultimately a variety of FAIR related tools, including software and hardware. It is GFF's intention to become the schema holder for community defined certifications. The actual certification process will be executed by a number of international certifying bodies. In this Pilot, GFF assists by providing prototype FAIR evaluation services for datasets via the Purple Polar Bear[26] a software development company located in the Netherlands.

---

[21] Science Europe Guidance Document, Presenting a Framework for Discipline-specific Research Data Management, January 2018.
https://www.scienceeurope.org/wp-content/uploads/2018/01/SE_Guidance_Document_RDMPs.pdf
[22] https://www.castoredc.com/
[23] https://github.com/DTL-FAIRData/FAIRDataPoint/wiki/FAIR-Data-Point-Specification
[24] Wilkinson, M. D. et al. A design framework and exemplar metrics for FAIRness. Sci. Data 5:180118 doi: 10.1038/sdata.2018.118 (2018); Evaluating FAIR-Compliance Through an Objective, Automated, Community-Governed Framework. Mark D Wilkinson, Michel Dumontier, Susanna-Assunta Sansone, Luiz Olavo Bonino da Silva Santos, Mario Prieto, Julian Gautier, Peter McQuilton, Derek Murphy, Merce Crosas, Erik Schultes bioRxiv 418376; doi: https://doi.org/10.1101/418376; http://linkeddata.systems:3000/FAIR_Evaluator/
[25] https://gofairfoundation.org

In this way, the science funders can require and monitor compliance with FAIR data-stewardship plans without the need for specialised expertise themselves. The evaluator service will not only run the FAIR metrics evaluations, but also deliver a certified report regarding compliance with the FAIR Principles and the resulting level of FAIRness. These certificates will be delivered to the science funders, ZonMW and HRB, as research outputs are published in the FAIR repositories. Hence, similar to the Data Stewardship plan, the funders will not have to actively review the FAIR compliance of the resulting research outputs, but they will instead receive trusted, external indications of FAIR compliance following community-defined certification schema.

---

[26] https://www.purplepolarbear.nl  Purple Polar Bear has already tested prototype evaluators in real-world FAIR hackathons, such as Bio-IT World in Boston, May 2018: http://www.bio-itworld.com/2018/06/06/finding-the-usable-in-fair-data-at-bioit18.aspx

## Pilot status and implementation

The FAIR Funders Pilot offers a solution to funders who wish to support research embedded within a FAIR Data Stewardship framework. As sketched here, the FAIR Funders Pilot will also be among the earliest multi-stakeholder implementations of the Internet approach to FAIR data and services. Presently, the technological components of the Pilot appear to be most advanced and ready for use (metadata authoring tools and registration services; simple and configurable data stewardship planning tools; FAIR data capture tools; FAIR repositories; FAIR metrics and evaluators). More challenging are the social components of the Pilot (organizational changes and incentives) but these too are also beginning to emerge (M4M workshops; early-mover funders that are clarifying their approach to FAIR research funding based on years of previous pilots; national initiatives for the development of FAIR certifications). Hence, this Pilot is poised to bring the necessary technological, administrative and social pieces in place to change fundamentally the way we do science.

The GO FAIR International Support and Coordination Office and the Research Data Alliance are well positioned to harness the substantial, ongoing activities of the initial stakeholders (and those stakeholder who are encouraged to join later) towards the Pilot objectives. In addition, numerous synergetic opportunities within the Pilot have been identified and will be developed by the Pilot participants very much like any other agile implementation project, for which the initial participants have ample expertise.

External to the Pilot, a public-private partnership of roughly 15 organizations known as the FAIR Service Provider Consortium (FSPC)[27] has recently emerged in response to the ever-increasing market demand for professional FAIR expertise, tooling and services. The FSPC follows the GO FAIR Rules of Engagement[28] with two key elements: commitment to FAIR Principles, and to a principle of 'no vendor lock in'. FSPC partners, Mobiquity and Phortos Consultants, have formed a partnership to service multiple projects relevant to the Pilot. For example, The Dutch Institute for Health Care (Zorg Instituut Nederland) has formally decided, based upon several earlier pilots, the GO FAIR implementation approach is recommendable for all Dutch Health Care Data. The FSPC is now preparing to embark on a Public-Private Partnership to turn all GO FAIR reference implementations into a professional online service suite, including each of the components described in the seven stages of the Pilot. Following the funders' interest in automated, professional and scalable suite of FAIR data services, the FSPC will be available to professionally extend the implementation results of the pilot, including robust hosting and maintenance of solutions for PI's, data stewards, and funders.

---

[27] https://osf.io/kpzfm/
[28] GO FAIR RoE https://www.go-fair.org/implementation-networks/how-to-become-an-implementation-network/rules-of-engagement/

# Appendix A

GO FAIR International Support and Coordination Office, Leiden, Netherlands
    E. A. Schultes[29] 0000-0001-8888-635X
    H. Pergl Sustkova 0000-0002-4462-6465
Max Planck Computing & Data Facility, Garching, Germany
    P. Wittenburg[30] 0000-0003-3538-0106
Health Research Board, Dublin, Ireland
    A. Montesanti 0000-0003-0413-2003
ZonMw, The Hague, Netherlands
    S. M. Bloemers 0000-0003-3710-3188
    S. H. de Waard 0000-0002-4691-7354
Stanford University, Stanford, CA, USA
    M. A. Musen 0000-0003-3325-793X
    J. B. Graybeal 0000-0001-6875-5360
Leiden University Libraries, Leiden, Netherlands
    K. M. Hettne 0000-0002-4182-7560
Leiden University Medical Center, Leiden, Netherlands
    A. Jacobsen 0000-0003-4818-2360
Czech Technical University in Prague, Faculty of Information Technology, Czech Republic
    R. Pergl 0000-0003-2980-4400
Dutch Techcentre for Life Sciences, Utrecht, Netherlands
    R. W. W. Hooft 0000-0001-6825-9439
    C. Staiger 0000-0002-6754-7647
    C. W. G. van Gelder 0000-0002-0223-2329
Castor EDC, Amsterdam, Netherlands
    S. L. Knijnenburg 0000-0002-2475-6254
Purple Polar Bear, Utrecht, Netherlands
    A.C. van Arkel 0000-0003-4668-2580
GO FAIR Foundation, Leiden, Netherlands
    B. Meerman 0000-0002-0071-2660
Centre for Plant Biotechnology and Genomics UPM – INIA, Universidad Politécnica de Madrid, Madrid, Spain
    M. D. Wilkinson 0000-0001-6960-357X
Oxford e-Research Centre (FAIRsharing), Department of Engineering Science, University of Oxford, Oxford, UK
    S-A Sansone 0000-0001-5306-5690
    P. Rocca-Serra 0000-0001-9853-5668
    P. McQuilton 0000-0003-2687-1982
    A. N. Gonzalez-Beltran 0000-0003-3499-8262
Australia's Academic and Research Network (AARNet), Sydney, Australia
    G. J. C. Aben 0000-0001-6925-4349
Universidade Federal do Estado do Rio de Janeiro, Rio de Janeiro, Brazil
    P. Henning[31] 0000-0003-0739-6442
    S. Alencar 0000-0002-2992-2215

---

[29] Pilot co-coordinator and corresponding author.
[30] Pilot co-coordinator.
[31] Also at Fundação Oswaldo Cruz, Rio de Janeiro, Brazil

   C. Ribeiro 0000-0002-9571-1707
   C. R. L. Silva 0000-0002-4327-6272
   L. Sayão 0000-0002-6970-0553
Instituto Brasileiro de Informação em Ciência e Tecnologia, Rio de Janeiro, Brazil
   L. Sales 0000-0002-3614-2356
Fundação Oswaldo Cruz, Rio de Janeiro, Brazil
   V. Veiga 0000-0001-8318-7912
   J. Lima 0000-0002-7560-1068
   S. Dib 0000-0001-9629-088X
   P. Xavier 0000-0002-1807-9963
   R. Murtinho 0000-0002-7224-8190
German Research Network - DFN, Berlin, Germany
   J. Tendel 0000-0001-8597-9868
Wageningen University and Research, Wageningen, Netherlands
   B. F. Schaap 0000-0003-2877-8597
   P. M. Brouwer 0000-0001-8183-0484
RIKILT, Wageningen University and Research, Wageningen, Netherlands
   A. K. Gavai 0000-0002-4738-190X
   Y. Bouzembrak 0000-0001-8028-0847
   H. J. P. Marvin 0000-0001-8603-5965
Phortos Consultants B.V., Waddinxveen, Netherlands
   A. Mons 0000-0001-8038-7572
VU University Amsterdam, Amsterdam, Netherlands
   T. Kuhn 0000-0002-1267-0234
Rijksmuseum, Amsterdam, Netherlands
   A. A. Gambardella 0000-0002-4930-2662
Maastricht University, Maastricht, Netherlands
   R. de Miranda Azevedo 0000-0002-7641-6446
Mobiquity, Amsterdam, Netherlands
   V. Muhonen 0000-0002-1812-5288
University Medical Center Utrecht, Utrecht, Netherlands
   M. van der Naald[32] 0000-0002-4949-625X
Netherlands Heart Institute, Utrecht, Netherlands
   N. W. Smit[33] 0000-0002-8924-4014
ORCID, Amsterdam, Netherlands
   M. J. Buys 0000-0001-7234-3684
NIOZ Royal Netherlands Institute for Sea Research, Texel, Netherlands
   T. F. de Bruin 0000-0001-9149-2095
Leiden University, Leiden, Netherlands
   F. Schoots 0000-0002-4385-9312
Consultant, Amsterdam, Netherlands
   H. J. E. Goodson 0000-0001-8056-2385
Chemistry Department, Imperial College London, UK
   H. S. Rzepa 0000-0002-8635-8390
Keith G Jeffery Consultants, Faringdon, UK
   K. G. Jeffery 0000-0003-4053-7825

---

[32] Also at preclinicaltrials.eu
[33] Also at preclinicaltrials.eu

Royal Holloway, University of London, London, UK
	H. P. Shanahan 0000-0003-1374-6015
Wiley, Hoboken, NJ, USA
	M. Axton 0000-0002-8042-4131
Science Data Software, Rockville, MD, USA
	V. Tkachenko 0000-0003-4265-235X
Jackson Laboratory, Farmington, CT, USA
	A. D. Maya 0000-0001-7951-3439
University of Notre Dame, Notre Dame, IN, USA
	N. K. Meyers 0000-0001-6441-6716
University of Florida, Gainesville, FL, USA
	M. Conlon 0000-0002-1304-8447